\documentclass[12pt]{iopart}
\usepackage{iopams}
\usepackage{amsfonts}
\usepackage{graphicx}
\usepackage{xcolor}
\usepackage{subfigure}
\usepackage{textcomp}
\usepackage{cite}
\usepackage{float}
\usepackage{soul}
\usepackage{stackrel}
\expandafter\let\csname equation*\endcsname\relax
\expandafter\let\csname endequation*\endcsname\relax
\usepackage{amsmath,amssymb}
\usepackage{esint}
\usepackage{color}
\usepackage[normalem]{ulem}
\begin{document}
\title[]{Fluctuation dominated phase ordering in coarse-grained depth models: Domain wall structures, extreme values and coarsening}
\author{Arghya Das and Mustansir Barma}
\address{TIFR Centre for Interdisciplinary Sciences, TIFR, Hyderabad 500046, India}


\begin{abstract}
\noindent
Models of particles driven by a one-dimensional fluctuating surface are known to exhibit fluctuation dominated phase ordering (FDPO), in which both the order and fluctuations appear on macroscopic scales. Highly dynamic and macroscopically broad interfacial regions, each composed of many domain walls, appear between macroscopically ordered regions and consequently the scaled correlation function violates the Porod law. We focus on two essential quantities which together quantify the unique characteristics of FDPO, namely the total number of domain walls and the length of the largest ordered domain. We present results in the context of coarse-grained depth (CD) models, both in steady state and while coarsening.
Analytic arguments supported by numerical simulations show that even though domain wall number fluctuations are very strong, the associated variance remains constant in time during coarsening. Further, the length of the largest cluster grows as a power law with multiplicative logarithms which involve both the time and system size. In addition, we identify corrections to the leading power law scaling in several quantities in the coarsening regime.
We also study a generalisation of the CD model in which the domain wall density is controlled by a fugacity and show that it maps on to the truncated inverse distance squared Ising (TIDSI) model. The generalised model shows a mixed order phase transition, with the regular CD model (which exhibits FDPO) corresponding to the critical point.
\end{abstract}

\section{Introduction} \label{introduction}

\begin{figure}[h]
\begin{center}
 \includegraphics[width=10cm,height=8cm]{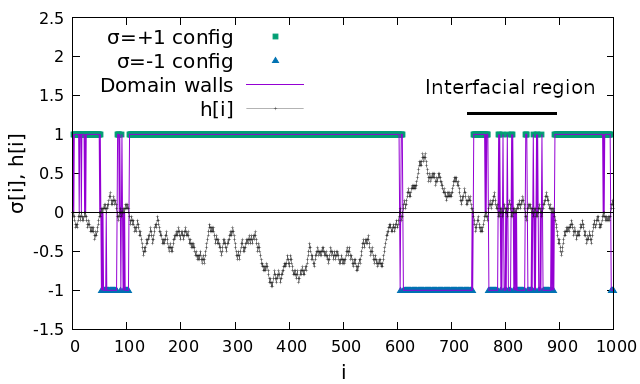}
\end{center}
\caption{A configuration of particles (spin $\sigma=+1$) and holes (spin $\sigma=-1$) in the CD model. There are stretches of pluses (green patches) and minuses (blue patches) separated by domain walls (vertical lines). Each domain wall is sharp, but a conglomeration of these forms an interfacial region. Note that the patches as well as the interfacial regions are broad.} \label{domain-wall}
\end{figure}

Fluctuation dominated phase ordering (FDPO) refers to a phase that exhibits the co-occurrence of the seemingly conflicting attributes of order and fluctuations, both at macroscopic scales. Macroscopic fluctuations of ordered regions occur in a variety of contexts, for example, particle systems driven by a fluctuating surface \cite{MB_fdpo, MB_LH, Shouri_FDPO}, collections of active particles and vibrating rods \cite{Giant-fluct-active}, nematic fluids \cite{Nematic_expt}, lipid clustering induced by actin dynamics on the cell membrane \cite{Aster_MRao}, granular systems \cite{granular_Rajesh}, and spin systems with truncated long-range interactions \cite{trunc_ising}.
Unlike in usual phase ordering, in FDPO a given ordered region dynamically breaks into or combines with other regions, leading to very large fluctuations. This leads to a state with several dynamic macroscopic ordered domains separated by domain wall structures which themselves are also macroscopically large (Figure \ref{domain-wall}). Existence of macroscopic ordered regions implies long-range order : the two-point correlation is a function of distance scaled by the system size, implying that in the thermodynamic limit the two-point correlation function has a nonvanishing large distance limit.
On the other hand, the broad domain wall structure leads to a breakdown of the Porod law \cite{Porod}, that is at small values of scaled distances the correlation function no longer decays linearly, but exhibits a distinctive cusp \cite{MB_fdpo}.

In this paper we characterise two crucial components of FDPO, namely the macroscopic ordered regions and the domain wall structures, within the coarse-grained depth (CD) model of a one-dimensional fluctuating surface \cite{MB_fdpo}. In the CD models, all heights above (below) a reference level are assigned spin values $\sigma=-1~(+1)$, and a cluster is defined as a stretch of like spins (Figure \ref{domain-wall}). The ordered regions in this model are formed by uninterrupted clusters of a single spin species, with no droplets of the other species.
Successive clusters are separated by domain walls, shown as vertical lines in Figure \ref{domain-wall}. The figure shows the occurrence of large interfacial regions composed of many domain walls, lying between macroscopically large clusters.
We focus on the behaviour of two quantities, namely the size $l_{\rm max}$ of the largest cluster and the total number of domain walls $S$, both in the steady state and during the approach to steady state. Interestingly, in 1D, the global quantity $S$ can be expressed as a sum of a local cross-correlation function \cite{MB_LH}.
The pair $(l_{\rm max},S)$ provides a compact and revealing characterisation of coarsening to FDPO as well as the FDPO steady state, which may be used as an alternative to a Fourier characterisation involving an infinite set of order parameters \cite{Kapri} in the CD model.

We study the evolution of $(l_{\rm max},S)$ during the approach to an ordered steady state governed by coarsening, as ordered domains are built up from a flat initial state. With time, several small patches of local order form, which then combine to form larger patches at larger times. The evolution is described in terms of a unique time dependent growing length scale $\mathcal{L}(t)$: at large distances and late times, the equal time two-point spin correlation function is a function of $r/\mathcal{L}(t)$, where $r$ is the separation of two spins and $\mathcal{L}(t)$ is defined as the coarsening length scale. $\mathcal{L}(t)$ typically grows as $t^{1/z}$ where $z$ is a dynamical exponent \cite{Bray94}, although, as discussed below, there may be significant corrections to the leading power law behaviour. Further, we expect that while coarsening, the locally ordered regions resemble the steady state, suggesting that the unusual nature of FDPO should be reflected during the time evolution as well.
The use of extreme value statistics to characterise $l_{\rm max}$ reveals that there are important multiplicative corrections to power law growth.
Further, because of the large fluctuations of the domains, one expects nontrivial domain wall structure to emerge during coarsening too.

In this paper we present the steady state and coarsening properties of $S$ and $l_{\rm max}$ in the CD models, and also show that the CD model lies at the separatrix of ordered and disordered phases. Below we outline the main highlights of the paper. 

\begin{itemize}
 
\item In the steady state the probability distribution of $S$ scales as $\sqrt{L}$, i.e. the spread in the number of domain walls is large, associated with a broad interface structure. On the other hand, the distribution of $l_{\rm max}$ scales with system size $L$ in the steady state, implying that both the mean and standard deviation are macroscopic, as one would expect in FDPO. We also identified interesting finite size corrections.

\item The behaviour of the coarsening length $\mathcal{L}(t)\sim t^{1/z}$ is asymptotically exact in the limit $t\rightarrow \infty$, which is taken after taking the thermodynamic limit $L\rightarrow \infty$. In general, at large but finite times, there may be significant corrections to the asymptotic power law. In the CD models, the correlation data is in fact consistent with $\mathcal{L}(t)\sim t^{1/z}\,(1+d/t^{\theta})$ at large times, where $d,\theta$ are constants. Further, both the number of domain walls $S$ and the largest cluster size $l_{\rm max}$ are affected by the corrections.

\item In the coarsening regime the density of domain walls carries a nontrivial time-dependence, $\langle S(t) \rangle \propto \sqrt{\mathcal{L}(t)}$. Remarkably, the variance of $S(t)$ is shown to be time-independent.
The largest cluster size $l_{\rm max}$ follows a Gumbel distribution, and the power law growth of $\langle l_{\rm max}(t)\rangle$ is modified by multiplicative logarithms in time and system size. The log corrections have a simple, generic origin in extreme value statistics, and should arise in other coarsening systems as well.

\item The CD models are related to other models, eg. the sliding particle (SP) models of particles driven by a growing surface \cite{MB_fdpo}, and the truncated inverse-distance squared Ising (TIDSI) model of long range interacting spins \cite{tidsi-2014}. In the former case, the CD model corresponds to the adiabatic limit of the SP model, where the surface evolves infinitely slowly allowing the particles to find the minimum energy configurations. In the other case, the equilibrium partition function of the TIDSI model at a particular parameter value corresponds to the steady state of generalised CD model discussed below.

\item We define a generalised CD (GCD) model by assigning a weight $\omega$ to each cluster in a CD configuration. A phase transition between ordered and disordered states is obtained on varying $\omega$. The regular CD model lies at the critical point $\omega_c=1$, which however is very different from a regular critical phase and exhibits FDPO behaviour. This is consistent with the observation that states exhibiting FDPO are often found at the separatrix of ordered and disordered phases, e.g. interacting particles on fluctuating surfaces \cite{MB_LH, Shouri_FDPO} and long-range spin systems \cite{trunc_ising}. The GCD model exhibits a mixed order transition, with a jump in the order parameter $\langle l_{\rm max} \rangle/L$, along with a divergence of the correlation length, as one approaches criticality from the disordered side.

\item Finally, the FDPO behaviour of the CD models can be compactly characterised using the pair $(l_{\rm max},S)$. This pair can also be used for the characterisation of the different phases of GCD models, and for other models in which ordered regions are associated with uninterrupted clusters of one species.
\end{itemize}

 This paper is organised as follows. In section 2 the different CD models, namely CD1, CD2, CD3 and CD4 models, are defined. Then in sections 3 and 4 we discuss simulation results for the CD3 model in the steady state and the coarsening regime respectively, and also outline the main results for the CD2 and CD4 models. In the next section (section 5) we define the generalised CD models and present the analytic results for the different phases and the nature of the phase transitions. We conclude in section 6.

\section{Coarse-grained depth (CD) models: Definition and correlation function}

\subsection{Definition of the CD models}

The CD models are defined as follows. Consider a fluctuating surface specified with a single-valued height field $h(x,t)$ defined on the one dimensional ring with coordinate $x$ at time $t$. The surface dynamics is governed by stochastic growth equations. Consider a {\it reference height} $h_r(t)$ intersecting the fluctuating surface. Follow this by a `coarse-graining' procedure: All points {\it below} the reference level are assigned a value $\sigma = +1$ while $\sigma = -1$ for all points above it; $\sigma = 0$ {\it at} the intersection points of the reference level and the surface. The configuration thus defined is a coarse-grained depth (CD) configuration consisting of domains of $+1$'s and $-1$'s of different widths. 

 An intuitive way to motivate the CD models is to imagine hard core particles relaxing to the local minimum of a random potential landscape that evolves in a time much larger than the relaxation timescale of the particles \cite{MB_fdpo}. The picture of domains becomes apparent if we imagine there are particles at all points with $\sigma=+1$. As the surface fluctuates in time, the domain widths change, and if we start with a flat surface at $t = 0$, we observe coarsening.

 We consider the surface dynamics corresponding to Edwards-Wilkinson (EW) and Kardar-Parisi-Zhang (KPZ) evolution in one dimension:
\begin{eqnarray}
 \frac{\partial h}{\partial t} &=& \frac{\partial^2 h}{\partial x^2} + \eta(x,t), ~~~~~~~~~~~~~~~~~~~{\rm (EW)}\nonumber \\
 \frac{\partial h}{\partial t} &=& \frac{\partial^2 h}{\partial x^2} +\lambda \left(\frac{\partial h}{\partial x} \right)^2 + \eta(x,t), ~~~{\rm (KPZ)}\nonumber
\end{eqnarray}
where $h(x,t)$ is the single-valued height field at position $x$ at time $t$, and $\eta(x,t)$ is Gaussian white noise. In the current work we study models defined on a 1D lattice. The dynamics is defined using the lattice versions of the EW and KPZ surface growth models. The lattice is periodic with $L$ sites; every site $i$ is assigned an integer height $h(i,t)$, and the height difference of any two neighbouring sites is defined as the `tilt' $\tau_{i+1/2}(t)=h(i+1,t)-h(i,t)$ with allowed values $\pm 1$. 
Since the height variable is single valued, the total tilt across the system is zero. The surface grows as local hills flip to become valleys, i.e. $\wedge \rightarrow \vee$, and vice-versa with certain rates. If these two rates are equal, the long-distance large-time behaviour is described by an EW dynamics, while unequal rates correspond to KPZ dynamics \cite{Daryaei2020, Barabasi-Stanley, Krug-Meakin92}.
\begin{figure}[h]
\begin{center}
 \includegraphics[width=10cm,height=7.5cm]{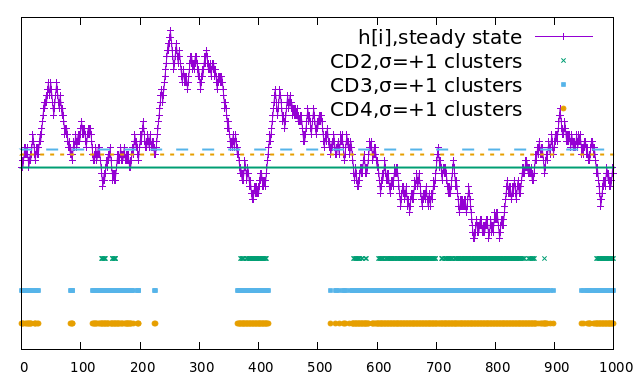}
\end{center}
\caption{Construction of spin configuration and domain structure in CD models. The fluctuating purple line $h(i)$ is an instanteneous height profile of a growing surface. The solid green line, blue dashed line, and red dotted lines cutting through the surface are the reference heights $h_r(t)$ for the CD2, CD3, and CD4 models respectively and the coloured points are the respective $+$ (up) spin domains. Domains of all sizes appear in the system.} \label{config}
\end{figure}

The reference level $h_r(t)$ plays a central role in defining different CD models. Assuming a flat surface at $t=0$ i.e. $h(x,0)=h_0$, and setting $h_r(t)=h_0$ defines the CD1 model; the CD2 model is defined by pinning the reference height to the height at $x=0$, i.e. $h_r(t)=h(0,t)$; for $h_r(t)=\frac{1}{L}\int_0^L h(x,t) dx$ which is the mean height of the surface, we have CD3 model; finally let us define a `Fermi level' $h_f(t)$ such that at every instant exactly half of the points are below $h_f(t)$, then the CD4 model is defined for $h_r(t)=h_f(t)$. The picture of relaxation of the particles in an infinitely slowly evolving potential landscape is most closely connected to the CD4 dynamics which conserves particle number. Placement of cuts which define different CD models is illustrated in Figure \ref{config} for a given surface height profile.
We shall not consider the CD1 model in the subsequent discussions, since at large times the surface diffuses away indefinitely far from $h_0$ rendering all spins to be parallel with unit probability in the steady state \cite{MB_fdpo}.

The steady state distributions of height profile $h_{\rm ss}(x)$ (and similarly $h_{\rm ss}(i)$) in both EW and KPZ models in a 1D spatial interval $L$ are identical to the distribution of a random walk trajectory \cite{Barabasi-Stanley}, with $x~({\rm or}~i)$ standing for time and the height difference for the displacement of the walker. In the present case the height field satisfies a periodic boundary condition, $h(i+L)=h(i)$. Thus, the corresponding random walk trajectory is constrained to return to the initial position after $L$ steps, which in the continuum limit reduces to a Brownian bridge. This identification helps in the numerical analysis of the state of the CD models since to generate a steady state height profile it is enough to draw the local tilts independently from a uniform distribution of tilts, provided the total tilt is zero. Given a height profile on a lattice, we do the coarse-graining of height fluctuations as discussed above.

\subsection{Two-point correlations}

In the steady state of the CD models there are always domains of size $\tilde{y}L$. The fraction $\tilde{y}$ fluctuates strongly giving rise to FDPO, which is manifest in the CD models both in steady state and while coarsening.
 \begin{figure}[h]
 \begin{center}
 \includegraphics[scale=0.45]{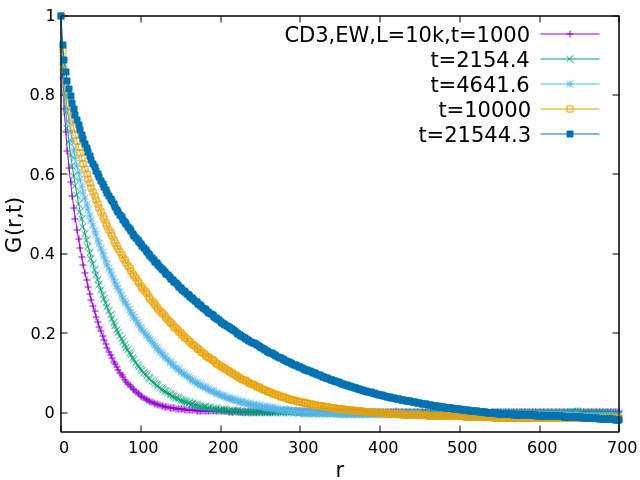} \hspace{0.2cm}
 \includegraphics[scale=0.45]{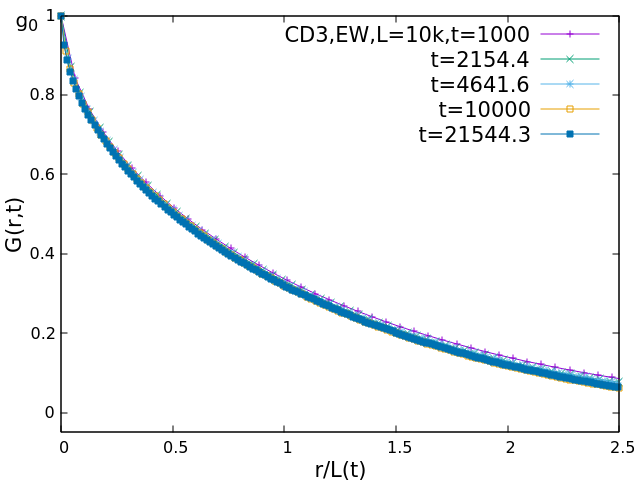}
 \end{center}
 \caption{ Left panel: Equal time two-point spin correlation $G(r,t):= \langle \sigma_i(t)\sigma_{i+r}(t) \rangle$ at different times for the CD3 model defined on a EW surface.
 Right panel: Scaling of the 2-point correlation and cusp at the origin, $G(r,t) = G(y=r/\mathcal{L}(t)) \approx g_0 - c\,y^\alpha$ for $y\rightarrow 0$ with $g_0=1,\alpha=0.5$. The scaling with $\mathcal{L}(t)\approx t^{1/z}$ signifies coarsening and that $\alpha$ is less than $1$ signifies the cusp exhibiting violation of the Porod law. Here $z=2$ is the dynamical growth exponent of the underlying EW surface. Similar behaviour is observed when the underlying surface is KPZ, except that in this case $z=3/2$.}\label{porod-violation}
 \end{figure}
Starting from a uniform initial profile, the domains start forming and gradually coarsen into larger domains characterised by the growing length scale $\mathcal{L}(t)$. 
The manifestation of coarsening to FDPO for the CD3 model is shown in Figure \ref{porod-violation} using the large-distance long-time scaling of equal time two-point correlation: $\lim_{r\rightarrow \infty}\,\lim_{\mathcal{L}(t)\rightarrow \infty} G(r,t)=\langle \sigma_i(t)\sigma_{i+r}(t) \rangle = G(y=r/\mathcal{L}(t))$ with $\mathcal{L}(t)\sim t^{1/z}$. The scaling function shows a cusp at small values of the scaled variable $y=r/\mathcal{L}(t)$, $G(y) \approx 1 - c\,y^{0.5}$ for $y\rightarrow 0$ implying violation of the Porod law. A more careful analysis shows that there are finite-time corrections to the leading power law form of $\mathcal{L}(t)$, which have important implications as discussed in section \ref{sec-finite-t-L}.

In section \ref{gcd-crit}, we define a generalisation of the CD model in which a weight $\omega$ is assigned to each cluster. It is shown that the generalised model exhibits a phase transition on varying $\omega$ and the regular CD model discussed above lies at the critical point separating the disordered and ordered phases.

\section{Steady state} \label{steady-state}

In this section we present the simulation results for the steady state properties of the number density of clusters $s=S/L$ and largest cluster size $l_{\rm max}$ in the CD models. Detailed results are given for the CD3 model, and results for other CD models are outlined at the end of the section.

\subsection{Properties of $s$ in the steady state}

\begin{figure}[h]
\begin{center}
 \includegraphics[width=10cm,angle=0.0]{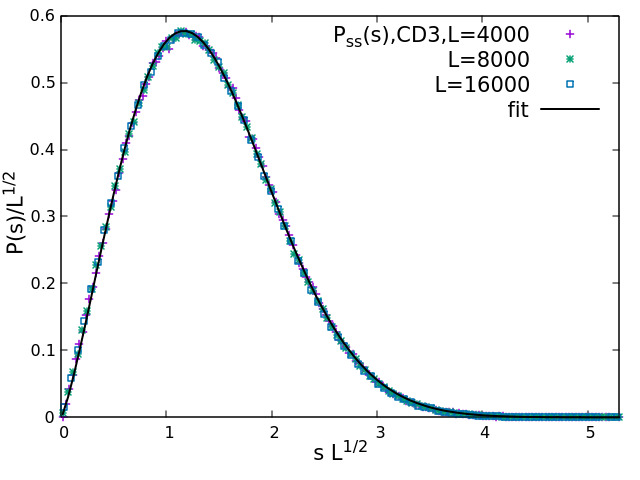}
\end{center}
\caption{System size scaling for the distribution of $s$ for the CD3 model in the steady state. The data are for system sizes $L=4000,8000,16000$. The scaled data fits well with $f(u)=0.89\,x^{1.25}\,e^{-0.46\,x^2}$ (Eq.~\eqref{ps_cd3ss}).} \label{Pss-s-cd3}
\end{figure}

Note that $s$ depends on the full configuration $\lbrace \sigma_i \rbrace$ and is therefore a global quantity. Following \cite{MB_LH} where two species of particles interact with a fluctuating surface and a {\it local} quantity is shown to capture much of the global phase behaviour, it is instructive to extend the approach to the CD models. To this end, we define the local cross-correlation between curvature (or `bend') $b(i,t)=\tau_{i+1/2}(t) - \tau_{i-1/2}(t)$ and spin $\sigma(i,t)$, 
\begin{equation}
 s(i,t) = b(i,t)\,\sigma(i,t),
\end{equation}
which can take one of the three values $+2,-2,0$ at each site. It turns out that $\sum_i s(i,t)$ gives the number of clusters, or equivalently the number of domain walls, in the CD models. One can see this as follows. Let $i_1$ and $i_2$ be the edge sites of a domain of up spins. In such a domain $s(i)$ takes a value $+2$ on a valley $\vee$, $-2$ on a hill $\wedge$, and $0$ on a slant edge. Therefore, the summation of $s(i)$ from ${i_1}$ to ${i_2}$ proceeds with consecutive cancellation of valleys and hills; however, since the domain of pluses has to cross over to a minus at the left of $i=i_1$ and at the right of $i=i_2$, i.e. the surface crosses the reference height before $i_1$ and $i_2$, there must be one excess valley compared to the hills in the domain. This is most easily seen in a domain of pluses where the surface profile contains only a single valley. Consequently, the aforementioned sum returns only a value $2$. The same argument goes through for a domain of minuses, where there is an extra hill, and this time each hill contributes $+2$ to the sum. Consequently, $S\equiv \frac{1}{2}\sum_{i=1}^L s(i)$ just gives the instantaneous total number of domain walls, or equivalently the number of points at which the reference level cuts through the surface, i.e. the number of domain walls.
Because of translational invariance, the statistical properties of $s(i)$ are independent of position, and we work with the density of domain walls $s=S/L$.

Recall that in the steady state the behaviour of 1D EW and KPZ surfaces are identical to the random walk trajectories. Consequently in the steady state of CD models $S$ is similar to the number of times a Brownian walker returns to the origin in an interval $L$ (the correspondence to return to origin is exact for the CD2 model). The average number of returns is $\sim \sqrt{L}$ \cite{Godreche}, and consequently, we expect,
\begin{equation}
 \langle s\rangle_{\rm ss}\equiv \frac{\langle S\rangle_{\rm ss}}{L}\sim L^{-1/2}. \label{s-ss}
\end{equation}
In fact, the full steady state distribution of $s$ scales as $1/\sqrt{L}$, as demonstrated in the Figure \ref{Pss-s-cd3}: 
\begin{equation}
 P_{\rm ss}(s)=\sqrt{L}\,f\left(s\,\sqrt{L}\right), \label{scaling-form}
\end{equation}
This implies that both the mean and standard deviation of the density $s$ of domain walls is of the same order ($\sim 1/\sqrt{L}$),
\begin{equation}
\sigma(s) \sim L^{-1/2}. \label{sigma-ss}
\end{equation}
Equivalently, the standard deviation $\sigma(S)$ in the total number of domain walls $S$, which equals $L\,\sigma(s)$, is proportional to $\sqrt{L}$, indicating large fluctuations in the domain wall structure. In the CD3 model, the scaling function $f(u)$ fits well to the following expression,
\begin{equation}
 f(u)\,=\,c_1\,u^{\beta}\,exp(-c_2\,u^2/2), \label{ps_cd3ss}
\end{equation}
where we found a good fit with, $\beta=1.25, c_1=0.89, c_2=0.92$ (Figure \ref{Pss-s-cd3}).
 
\subsection{Statistics of the largest cluster $l_{\rm max}$}
 
\begin{figure}[h]
 \begin{center}
 \includegraphics[scale=0.43]{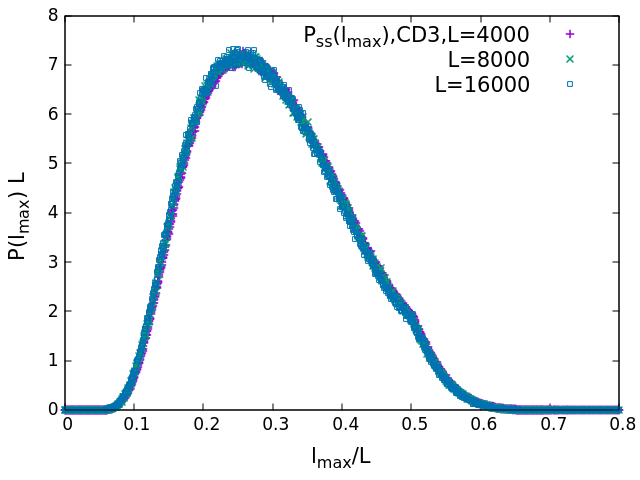}\hspace{0.2cm}
 \includegraphics[height=5.5cm,width=7.6cm]{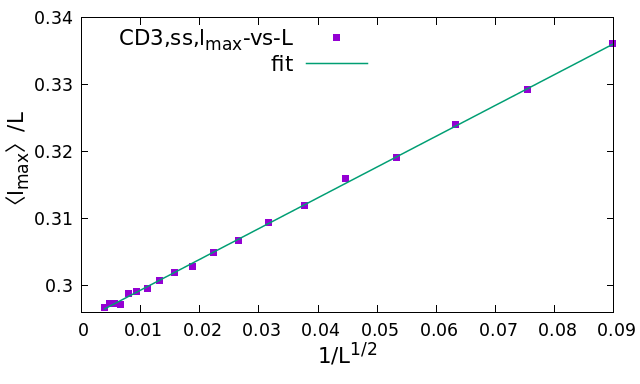}
 \end{center}
 \caption{Left panel: steady state distribution of the largest cluster size in the CD3 model, scaled with the system size. The scaling indicates that not only the average but the standard deviation and roots of all higher cumulants are macroscopic. Right panel: Simulation results for the finite size effect in the average largest cluster size in the steady state. The data for $\langle l_{\rm max} \rangle$ is scaled by system size $L$ and plotted against $1/\sqrt{L}$, which indeed fits well to a straight line $a_0+b_0\,x$ in accordance with Eq.~\eqref{lm-ss}. The data indicates, $a_0=0.2947,b_0=0.46$ for CD3.} \label{lmax-ss-cd3}
 \end{figure}
While $s$ is a measure of the density of the domain walls, the ordering is characterised by the size of the clusters. Recall that to characterise macroscopic order in these models, we have to look at the maximum cluster size $l_{\rm max}$, that is the size of the largest of the $S \sim \sqrt{L}$ number of domains. In the context of FDPO, the steady state behaviour of the largest cluster size was studied earlier for the sliding particle (SP) models and it was found numerically that both the mean and standard deviation of $l_{\rm max}$ are of the same order, growing as a sublinear power of the system size \cite{Sakuntala2006}.
In the CD models, because of the mapping of the steady state surface profiles to the trajectories of a random walker, each CD cluster is isomorphic to the first return time of the walker. Consequently we find that the cluster sizes $`l'$ are distributed as ${\rm Prob.}(l) \sim l^{-3/2}$ with a cut-off \cite{MB_fdpo}. However the domain sizes are not independent as the total tilt is zero due to the periodic boundary condition. We find numerically that $\langle l_{\rm max} \rangle_{\rm ss} \propto L$ signifying the emergence of macroscopic order. 
In fact, the entire distribution and thus the standard deviation of $l_{\rm max}$ scales with system size $L$, implying macroscopic fluctuations as expected for FDPO (left panel of Figure \ref{lmax-ss-cd3}). The distribution itself is nontrivial and the data shows it to be a skewed distribution with a cusp (break of slope) at $l_{\rm max}=L/2$, as also observed in the extremal cluster statistics of other one-dimensional conserved systems \cite{Krapivsky95, Barkai2020, Godreche-extr}.

Beyond the asymptotic $(L\rightarrow \infty)$ form of $P_L(l_{\rm max})$, we expect to see finite size corrections. Indeed, such corrections are seen prominently in the mean value of the largest cluster $\langle l_{\rm max} \rangle_{\rm ss}$, which shows a correction of the form in Eq. \eqref{lm-ss}
(right panel of Figure \ref{lmax-ss-cd3}),
\begin{equation}
 \langle l_{\rm max} \rangle_{\rm ss} = L\,\left(a_0 + \frac{b_0}{\sqrt{L}} \right). \label{lm-ss}
\end{equation}
The correction term in Eq. \eqref{lm-ss} is reminiscent of the behaviour of the longest excusrion of a Brownian walker in time window $[0,T]$ studied in \cite{Schehr-excursion}. If $\tilde{\tau}_{\rm max}$ is the longest time interval in which the walker does not cross the origin, the mean $\langle \tilde{\tau}_{\rm max} \rangle$ is proportional to $T$ to the leading order, and carries a subleading correction of $O(\sqrt{T})$. In view of the close correspondence of the CD clusters in steady state to the first return time of a random walk, this is consistent with Eq. \eqref{lm-ss}.
However this correspondence is not precise, as the excursions considered in \cite{Schehr-excursion} are unconstrained, whereas in the CD models the surface is always closed, which corresponds to trajectories returning to the origin.

\begin{figure}[h]
 \begin{center}
 \includegraphics[scale=0.45]{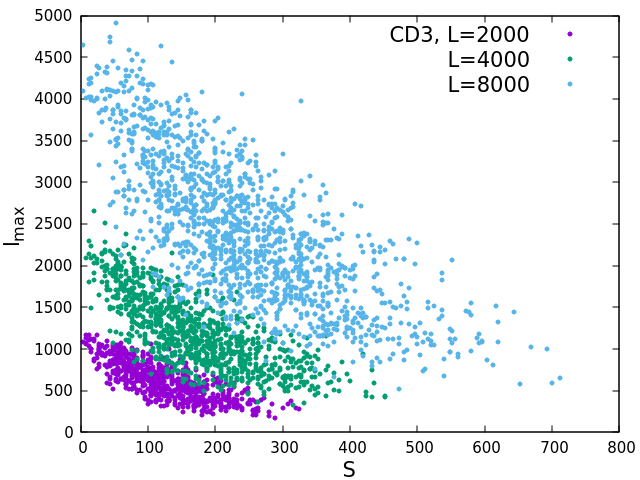}\hspace{0.2cm}
 \includegraphics[scale=0.45]{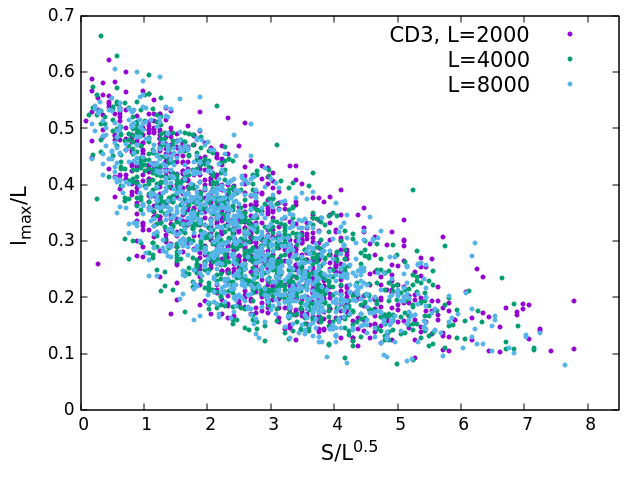} 
 \end{center}
 \caption{Left panel: Scatter plot of largest cluster size $l_{\rm max}$ and the number of clusters $S$ in the steady state of CD3 model for system sizes $L=2000,4000,8000$. The data indicates strong anti-correlation between these two quantities. In the right panel the scaled plot $l_{\rm max}/L$ vs $S/\sqrt{L}$ is shown.} \label{scatter-cd3}
 \end{figure}

 It is interesting to ask about the correlation between $l_{\rm max}$ and $S$. In the CD models the clusters are distributed as $l^{-3/2}$, implying that clusters of all sizes including macroscopic clusters occur. One expects that in the configurations having one very large cluster the number of clusters will be generically low. Indeed the scatter plot of $l_{\rm max}$ vs $S$ shown in the left panel of Figure \ref{scatter-cd3} indicates a strong anti-correlation between these two quantities. The right panel shows the scatter plot data exhibiting scaling.

\subsection{Steady state in other CD models}

For all three CD models, the full steady state distribution of $s$ satisfies the general scaling form,
\begin{equation}
 P_{\rm ss}^{j}(s)=\sqrt{L}\,f^{(j)}(s\,\sqrt{L}\,),
\end{equation}
analogous to Eq. \eqref{scaling-form}; here $j=2,3,4$ stands for the different CD models. The scaling function $f^{(j)}(u)$ is given by Eq. \eqref{ps_cd3ss}, where the constants $(\beta,c_1,c_2)$ take values $(1,1,1)$ for the CD2 model \cite{Godreche} and $(2,6.383,4)$ for the CD4 model (numerically obtained).
Note the differences in the small $u$ behaviour of the three CD models, that gives the probability of having very small number of clusters in the system: in the CD2 model $f^{(2)}(u)\sim u$ for $u\rightarrow 0$. In this model there are several configurations where the entire surface lies above the reference level, corresponding to only one cluster. In the CD3 model, however, the reference level always cuts the surface and therefore a very small number of clusters is less likely (reflected in $f^{(3)}(u)\sim u^{1.25}$ for small $u$). In the CD4 model, because the total number of spins in each species are conserved and equal, the cluster number tends to be much higher and the probability $f^{(4)}(u)$ falls even faster, as $u^2$ as $u\rightarrow 0$.
These results are valid for both EW and KPZ surfaces which have identical steady states in one dimension.

 In all the CD models the distribution of the largest cluster size $l_{\rm max}$ scales with system size $L$, signifying macroscopic order and equally large fluctuations. In this case the distributions are entirely different in the three models. Because of the correspondence of the steady state of the CD2 model to the Brownian bridge, the distribution of the largest cluster size is known analytically \cite{Godreche}. In the CD3 and CD4 models however we do not have analytical predictions for the form of the distribution. It is observed numerically that, in the CD4 model the distribution of $l_{\rm max}$ shows strong `even-odd' effects. Despite these differences, the average value of the largest cluster sizes exhibit finite size effects given by Eq. \eqref{lm-ss} in all the CD models, with the constants $(a_0,b_0)$ taking different values.

\section{Coarsening towards FDPO} \label{coarsening-cds}

We start with a macroscopically flat profile for the surface height, realised on a 1D lattice by choosing $h[i,0]=\mod(i,2)$, and let the surface evolve with EW or KPZ dynamics. In the initial state, we have $s(0)=1, l_{\rm max}(0)=1$. As the surface coarsens and roughens in time, longer domains of the CD model form, and consequently $s(t)$ decreases while $l_{\rm max}(t)$ increases. In an infinite system undergoing coarsening, at late times $t$, a characteristic coarsening length $\mathcal{L}(t)$ emerges within which there are strong correlations, and beyond which correlations are weak. In the CD3 model a study of correlation functions (Figure \ref{porod-violation}) shows that to leading order, we have $\mathcal{L}(t)\sim t^{1/z}$ with $z=2$ for EW and $z=3/2$ for KPZ dynamics \cite{MB_fdpo}.

\subsection{Corrections to power law scaling in $\mathcal{L}(t)$}

 Strictly speaking, the $t^{1/z}$ behaviour for the coarsening length scale is only the leading behaviour in time $t$, and would hold as $t\rightarrow \infty$; one may have power law corrections to the leading behaviour. For example, within the renormalisation group (RG) theory of critical phenomena, scaling behaviour characterised by a single power law occurs very close to the fixed point along the {\it relevant directions} only. If the system is slightly away from the fixed point, confluent corrections to scaling can arise from irrelevant operators at the corresponding fixed point \cite{Wegner72, Jasnow75, Aharony83}. These generate significant power law corrections to the asymptotic decay of the static correlation functions as well as nontrivial dependence on parameters (eg. the reduced temperature) \cite{Brezin,Brezin-prl}.

\begin{figure}[h]
 \begin{center}
 \includegraphics[height=4.3cm,width=5.1cm]{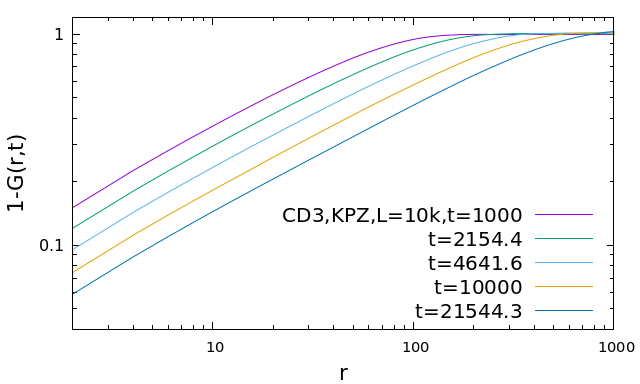} 
 \includegraphics[height=4.5cm,width=5.1cm]{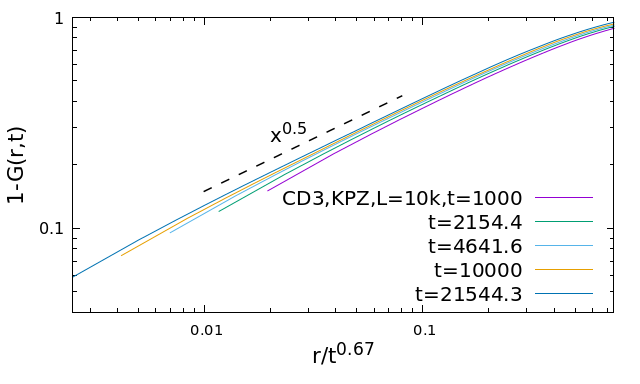} 
 \includegraphics[height=4.3cm,width=5.1cm]{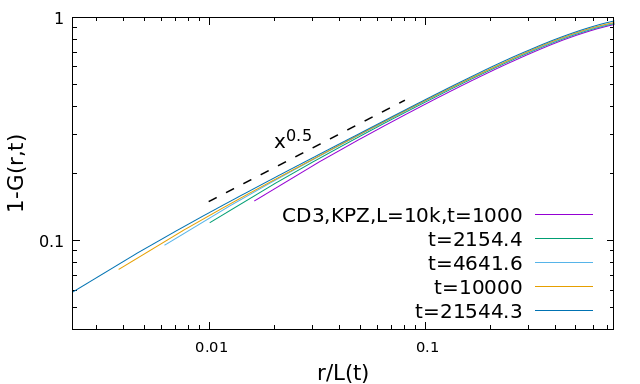}
 \end{center}
 \caption{Left panel: Equal time two-point correlation $(1-G(r,t))$ at different times for the CD3 model defined on a KPZ surface. Middle panel: Attempted scaling of the correlation function with $t^{1/z}$, where $z=3/2$.
 Right panel: Scaling of the correlation with $\mathcal{L}(t)$ in Eq. \eqref{lm-finite-t}, which shows an improved data collapse with $d=2$. Here $\theta$ is $1/(2z)=1/3$.
 The dashed lines in the middle and right panels show the `cusp' exponent to be $0.5$, implying strong deviation from the Porod law.}\label{Lt-correction}
 \end{figure}

 For critical dynamics, similar effects can give rise to corrections to dynamic scaling as well \cite{Hohenberg77}. In the present case $t$ being large but finite
can lead to significant corrections to the power laws. Indeed, our data can be explained by a leading power law correction to the coarsening length scale,
\begin{equation}
 \mathcal{L}(t) \approx C\, t^{1/z}\left(1+\frac{d}{t^\theta}\right), \label{lm-finite-t}
\end{equation}
$C,d,\theta$ are constants. Arguments given later in the section suggest that $\theta=1/(2z)$. If the finite time correction is ignored and the data is `force-fitted' to the form, $\mathcal{L}(t)\sim t^{1/z'}$, then the effective exponent $z'$ differs from its value for the underlying surface growth models. Revisiting the scaling of equal-time two point correlation in the CD3 model, we find significantly better data collapse when corrections are included.
In Figure \ref{Lt-correction} the comparison of the correlation data scaled with $\mathcal{L}(t)$ without and with the correction term is shown for the CD3 model defined on the KPZ surface and improved scaling is observed with $d\simeq 2$. This also clearly brings out the `cusp' near for smaller values of $r/\mathcal{L}(t)$, with cusp exponent $0.5$.
The correction in $\mathcal{L}(t)$ also has strong effects on the behaviour of domain wall density, cluster statistics and largest cluster size, all of which show significant deviations from a single power law in time. This is discussed below and also shown in Figs. \ref{mean-std-st-cd3} and \ref{cluster-coarsening}.
For the model defined on the EW surface the correction term turns out to be relatively less prominent and we found $d\simeq 0.2$ (not shown). We have set $C=1$ in the figures in which $\mathcal{L}(t)$ is used.

 For extremal quantities such as $l_{\rm max}$, there is another simple but significant effect which gives rise to multiplicative logarithms to the leading order. To see this, consider the following description. (i) At time $t$, a large system of size $L$ may be considered to be composed of subsystems of length $\mathcal{L}(t)$. The number of subsystems is $N_{\rm sub}(t) = L/\mathcal{L}(t)$, and we assume that they are statistically independent. (ii) Statistical properties at the scale of the subsystem (on length scales $\mathcal{L}(t)$) resemble those in the steady state of the full system (on length scales $L$). Thus characteristics of FDPO in the steady state of CD models should be reflected in the coarsening system on the scale of $\mathcal{L}(t)$.

 This picture leads to a couple of striking predictions: (i) The variance of $s(t)$ remains constant in time, although the mean value decreases as $\mathcal{L}(t)^{-1/2}$. (ii) The maximum cluster size $l_{\rm max}(t)$ grows as $\mathcal{L}(t) \log (N_{\rm sub}(t))$. The origin of the multiplicative logarithm rests on an argument of general validity, and should hold in other coarsening systems as well.

 The predictions discussed above and their validation by numerical simulation are presented below.

\subsection{$s$ in the coarsening regime}

\begin{figure}[h]
\begin{center}
  \includegraphics[width=10cm,height=6.5cm]{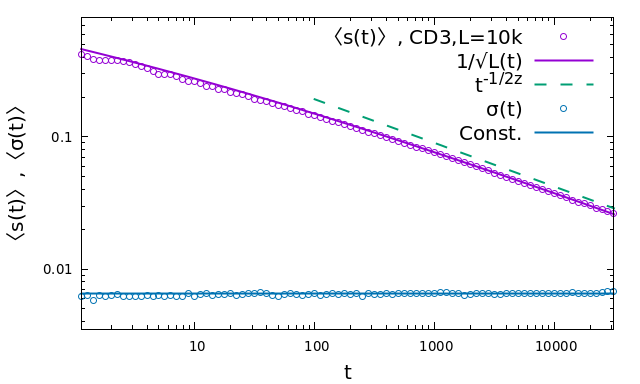}
 \end{center}
\caption{The mean and standard deviation of $s$ are plotted with time for the CD3 model defined on a KPZ surface. Here the system size is $L=10000$. The mean of $s(t)$ decays with time as $\langle s(t) \rangle \sim 0.8/\sqrt{\mathcal{L}(t)}$ which is the expected behaviour, with $\mathcal{L}(t)$ given by Eq. \eqref{lm-finite-t}. The standard deviation $\sigma(t)$ remains constant at a value $\approx 0.0065$. The green dashed line shows the behaviour without the power law correction in $\mathcal{L}(t)$, with $z=3/2$.} \label{mean-std-st-cd3}
\end{figure}

The properties of $s$ are obtained as follows,

\begin{equation}
\langle s(t) \rangle = \frac{\langle S(t) \rangle}{L} = \frac{\langle S_{\rm sub}(t) \rangle\, N_{\rm sub}(t)}{L} = \frac{\langle S_{\rm sub}(t) \rangle}{\mathcal{L}(t)} \propto \mathcal{L}(t)^{-1/2} = t^{-1/2z}+{\rm corrections}, \label{avs-t}
\end{equation}
where $S_{\rm sub}(t)$ is the value of $S$ evaluated in a subsystem of size $\mathcal{L}(t)$ at a time $t$, and at the last step we have used $\langle S_{\rm sub} (t)\rangle \sim \sqrt{\mathcal{L}(t)}$ which follows from Eq. \eqref{s-ss} using the local steady state argument.
Numerical results for $\langle s(t) \rangle$ are consistent with Eq. \eqref{avs-t} (Figure \ref{mean-std-st-cd3}).

 Next, we argue that the standard deviation of $s$, namely $\sigma(t)$, is constant in time. Define $Var(S,t)\equiv L^2\,\sigma^2(t)$ as the variance of $S$ in the entire system, and $Var_{\rm sub}(S,t)$ as the variance of $S$ in a subsystem. Then, following the steps as in Eq. \ref{avs-t},
\begin{equation}
 \sigma^2(t)=\frac{1}{L}\frac{Var(S,t)}{L} \approx \frac{1}{L}\frac{Var_{\rm sub}(S,t)\,N_{\rm sub}(t)}{L}=\frac{1}{L}\frac{Var_{\rm sub}(S,t)}{\mathcal{L}(t)}={\rm const}\,\frac{1}{L}, \label{vars-t}
\end{equation}
where we have used the independence of the subsystems to write $Var(S,t)$ as $N_{\rm sub}(t) \times Var_{\rm sub}(S,t)$. Now, from Eq. \eqref{sigma-ss} we find that the variance of $s$, namely $Var(s)$, is proportional to $L^{-1}$ in the steady state, implying that the $Var(S)=L^2\,Var(s)$ is of $\mathcal{O}(L)$. The local steady state then suggests that, the variance $Var_{\rm sub}(S,t)$ of $S$ in a subsystem of size $\mathcal{L}(t)$ is proportional to $\mathcal{L}(t)$ itself, which gives the last equality in Eq. \eqref{vars-t}, and explains the time independence of $\sigma(t)$. Numerical results support the constancy of $\sigma(t)$, as shown in Figure \ref{mean-std-st-cd3}.

\subsection{Behaviour of $l_{\rm max}(t)$ during coarsening}

\begin{figure}[h]
 \begin{center}
 \includegraphics[height=6cm,width=7.5cm]{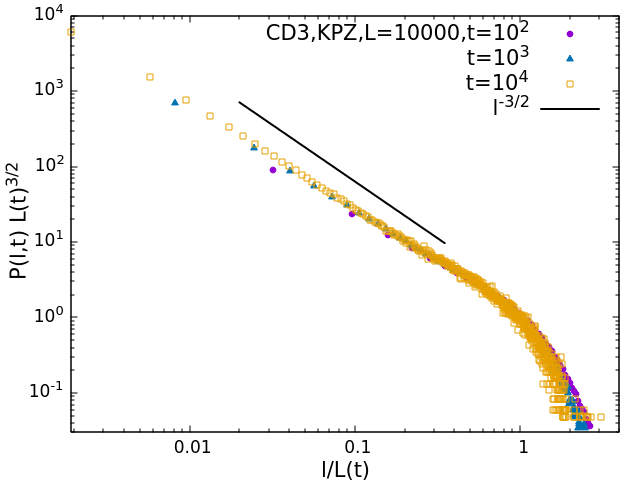}\hspace{0.3cm}
 \includegraphics[height=6cm,width=7.5cm]{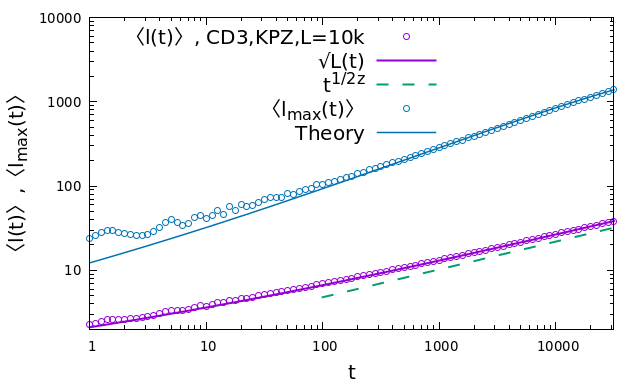}
 \end{center}
 \caption{Left panel: Cluster size distribution scaled with $\mathcal{L}(t)$ (Eq. \eqref{lm-finite-t}) at different times for CD3 model defined on a KPZ surface. The distribution is $P(l)\sim l^{-3/2}$ with a cut-off at a sizes $\sim \mathcal{L}(t)$ implying that, $\langle l(t) \rangle \sim \sqrt{\mathcal{L}(t)}$. In the right panel the time evolution of the average cluster size and average maximum cluster size for CD3 model on a KPZ surface is shown. Note that, the average cluster size indeed grows as $\sqrt{\mathcal{L}(t)}$ where ${\mathcal{L}(t)}$ includes the correction term given in Eq. \eqref{lm-finite-t}. The single power law behaviour in the absence of the correction term is shown by the green dashed line (with $z=3/2$). On the other hand the largest cluster size $\langle l_{\rm max}(t) \rangle$ evolves very differently from $\mathcal{L}(t)$ and contains logarithmic effects in time and system size (Eq. \eqref{lm-finite-tL}).} \label{cluster-coarsening}
\end{figure}

We have seen that ordering in the steady state can be characterised by $l_{\rm max}$. In the coarsening regime also, clusters of all sizes appear in the system and the distribution $P(l,t)$ of cluster sizes $l$ is given by a power law $\sim l^{-3/2}$ with a time dependent `cut-off' at length scales $\mathcal{L}(t)(\approx t^{1/z})$. Consequently, the average cluster size grows as $\langle l(t)\rangle \sim \sqrt{\mathcal{L}(t)}$, which is much smaller compared to the coarsening length scale. Therefore, to characterise ordering in the coarsening regime, we study the time dependent properties of the  global maximum cluster size $l_{\rm max}(t)$. Interestingly, it turns out that $\langle l_{\rm max}(t)\rangle$, is \emph{not} given by $\mathcal{L}(t)$ (Figure \ref{cluster-coarsening}) and the numerically available data cannot be fit by a power law across the full range. As we shall see, the finite system size and finite time effects strongly affect even the leading order behaviour of $\langle l_{\rm max}(t) \rangle$ through multiplicative logarithms.

\subsubsection{Finite time and finite size effect: Logarithmic behaviour} \label{sec-finite-t-L}
~\\
Let us investigate the time dependent behaviour of $l_{\rm max}(t)$.
We incorporate the effect of finite system size $L$ as follows.
As shown in the left panel of Figure \ref{cluster-coarsening}, the cluster size distribution $P(l,t)$ is a power law with cut-off at length scales $\sim \mathcal{L}(t)$, where for finite times the value of the length scale is given by Eq. \eqref{lm-finite-t}.
The local steady state argument suggests that there are many clusters that are {\it locally macroscopic} at the scale $\mathcal{L}(t)$ of the subsystems and one can intuitively associate the ordering length scale with the average size of all such clusters. The globally largest cluster is the maximum of all these locally macroscopic clusters.
Since the system is homogeneous and the subsystems are almost independent, we expect the cluster size distribution within a subsystem to be given by the cluster size distribution $P(l,t)$ as in the full system. In particular at scales $l\sim \mathcal{L}(t)$ there is a tail that decays rapidly : numerical results suggest that in the CD3 model the decay is exponential; consequently, the locally largest cluster size is sampled from that exponential tail.

\begin{figure}[h]
\begin{center}
 \includegraphics[height=6cm,width=7.6cm]{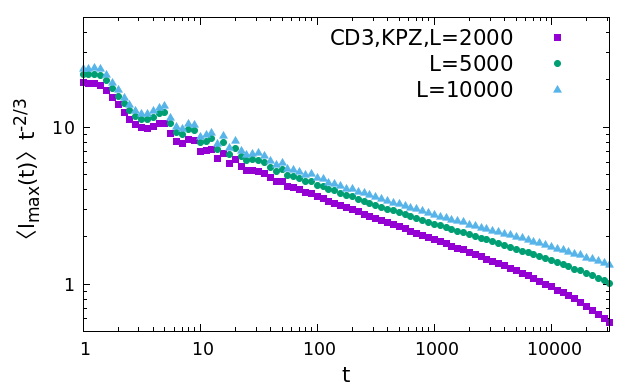}\hspace{0.2cm}
 \includegraphics[height=6cm,width=7.5cm]{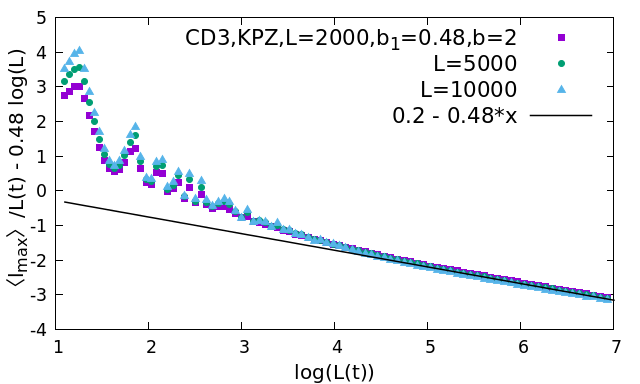}
\end{center}
\caption{System size and finite time effects in the average maximum cluster size in CD3 model defined on a KPZ surface. In the left panel we have shown the data for $\langle l_{\rm max}(t) \rangle/t^{2/3}$ vs $t$ for $L=2000,5000,10000$, showing strong $L$ and $t$ dependent departures from simple power law behaviour. In the right panel we plotted the data as $\left(l_{\rm max}/\mathcal{L}(t) - 0.48\,\ln L\right)$ vs $\ln \mathcal{L}(t)$ with $\mathcal{L}(t)$ taken from Eq. \eqref{lm-finite-t} and setting $C=1$. The observed data collapse is in agreement with Eq. \eqref{lm-finite-tL} with parameters, $a_1'=a_1\simeq 0.2,g_1\simeq 0.48, d\simeq 2, \theta=1/3.$} \label{lm-tL-cd3-kpz}
\end{figure}

To evaluate the properties of the globally largest cluster, we then have to look for the extremal statistics over the set of largest local clusters within a subsystem. Now the number of subsystems is $N_{\rm sub}(t)=L/\mathcal{L}(t) \gg 1$ in the coarsening regime, each contributing a local maximum cluster: the globally largest cluster is just the largest of all these local maxima. Therefore, in the CD3 model, the problem of finding the distribution of the globally largest cluster size reduces to the evaluation of the extremal properties of a number $N_{\rm sub} (t)$ of samples drawn from an exponential: assuming these samples to be statistically almost independent, at large $l$, $P(l,t)\sim e^{-w},~ {\rm with}~ w=l/(C_1\,\mathcal{L}(t)),~ {\rm where}~C_1$ is an unknown constant. The largest of $N_{\rm sub}(t)$ samples drawn from the distribution $e^{-w}$ is known to follow the Gumbel distribution\cite{Gumbel,Arnab-extreme} with mean $\langle w_{\rm max} \rangle = \gamma + \ln N_{\rm sub}(t), ~ {\rm where}~ \gamma$ is the Euler constant. Consequently, $\langle l_{\rm max} \rangle = C_1\,\mathcal{L}(t)\,\langle w_{\rm max} \rangle$ assumes the form,
\begin{eqnarray}
 \langle l_{\rm max}(t) \rangle &\approx & \mathcal{L}(t)\left[a_1 + g_1 \log(N_{\rm sub}(t)) \right] = \mathcal{L}(t)\left[a_1 + g_1 \log(L/\mathcal{L}(t)) \right]\\
 &\approx & C\,t^{1/z}\left(1+\frac{d}{t^\theta}\right)\left[a_1' + g_1\,\log(L) - g_1\,\log \lbrace t^{1/z}\,(1+d/t^\theta\,)\rbrace \right]
 \label{lm-finite-tL}
\end{eqnarray} 
where $a_1,g_1$ are constants and $a_1'=a_1-g_1\,\ln C$. At very large times the correction term $d/t^{\theta}$ can be neglected in the last expression, but the multiplicative logarithms remain, and $\langle l_{\rm max}(t) \rangle \approx C\,t^{1/z}\left[a_1' + g_1\,\log(L) - g_1/z\,\log(t) \right]$ to the leading order. Eq. \eqref{lm-finite-tL} predicts strong finite size and finite time effects in the maximum cluster size during coarsening. Using the value of $d$ obtained from the two-point correlation, Eq. \eqref{lm-finite-tL} is in very good agreement with the data for $\langle l_{\rm max}(t) \rangle$ at larger times, shown in Figure \ref{lm-tL-cd3-kpz}.

Note that the correction term in $\mathcal{L}(t)$ (Eq. \eqref{lm-finite-t}) contributes to the leading correction in time in Eq. \eqref{lm-finite-tL}. In the $t\rightarrow \infty$ limit, we expect to recover the steady state results with the replacement $\mathcal{L}(t)\rightarrow L$, or $t \rightarrow a\,L^z$, where $a$ is a constant. Using this in Eq. \eqref{lm-finite-tL} gives,
\begin{equation}
 \langle l_{\rm max} \rangle_{\rm ss} \approx a\,C\, L \left[(a_1' - g_1\,\log a) + (a_1' - g_1\,\log a -g_1)\frac{b\,a^{-\theta}}{L^{z\,\theta}} + \mathcal{O}\left(L^{-2\,z\,\theta}\right)\right]. \label{lm-ss-limit}
\end{equation} 
Comparing this with the finite size correction term of $\langle l_{\rm max} \rangle_{\rm ss}$ in Eq. \eqref{lmax-ss-cd3}, one obtains $z\theta=1/2 \Rightarrow \theta=1/(2z)$, which equals $1/3$ for KPZ and $1/4$ for EW.

A similar method of dividing the system into temporally growing subsystems was employed earlier in a different context, namely for finding maxium relative height in surface growth models \cite{Comtet2005}. The approach taken in this section to find the extremal cluster size is quite general and should be applicable to systems coarsening to an ordered state. Thus extremal behaviour during coarsening should exhibit nontrivial dependence (typically, multiplicative logarithms) on the system size and time. It would be interesting to check this for extremal properties of coarsening systems in general.

\subsection{Coarsening in other CD models}
The coarsening behaviour of the CD2 and CD4 models is analogous to the CD3 model. Using the scaling analysis of two point correlations, we found that in all the CD models $\mathcal{L}(t)$ is proportional to $t^{1/z}$ to the leading order, modified by the corrections as in Eq. \eqref{lm-finite-t}, where $z$ is the dynamical exponent of the underlying fluctuating surface.

 In all the CD models $\langle s(t)\rangle$ falls approximately as $t^{-1/2z}$ to the leading order, while the variance remains time independent in the CD3 and CD4 models. These follow from the local steady state argument discussed earlier. However, for CD2, we observe that the variance is proportional to $\langle s(t) \rangle^2$. A closer look reveals that since in CD2 the reference level is set at the height of a particular site (here site $0$), a single flip of the site can cause a global change in the number of cuts and hence the fluctuations in $s$ and dominated by the height fluctuation of the site. It corroborates the observation that the distribution $P(s,t)$ for CD2 has a bimodal nature, while that for CD3 and CD4 are unimodal (not shown); this qualitatively accounts for larger and time dependent fluctuations and the apparent violation of local steady state behaviour in CD2.

 The largest cluster size $l_{\rm max}(t)$ on the other hand shows strong deviation from the simple power law behaviour in all the CD models and in fact possesses nontrivial $L$ and $t$ dependence. However we found that, unlike in CD3, $\langle l_{\rm max}(t)\rangle$ in the CD2 and CD4 models cannot be explained well with the simple multiplicative logarithm $\ln(N_{\rm sub}(t))$. This is possibly related to the non-exponential form for the cut-off in the cluster size distribution $P(l,t)$. We do not consider this question further in the present paper.

\section{Generalised CD models: Phase diagram} \label{gcd-crit}
In this section we define a generalisation of the CD model and study the behaviour in different phases and at criticality.

 Recall that, in the steady state the height profiles of the EW and KPZ surfaces in one dimension are described by simple random walk trajectories and consequently all configurations appear with equal weights. A natural description of the corresponding CD configurations is given in terms of the cluster lengths: $\mathcal{C}=\lbrace l_1,l_2,\cdots l_{S};S | \sum_{k=1}^{S} l_k=L \rbrace$, $l_k$ is the length of the $k$'th cluster and $S$ is the number of clusters in that configuration. Here a cluster is defined as an unbroken stretch of pluses or minuses; in the configurations where the reference height passes through the sites and therefore there are sites having $\sigma = 0$ (as in CD2), a cluster is defined by the number of bonds between two successive $\sigma = 0$'s. A given configuration of clusters correspond to many possible configurations of the surface heights.

 We generalise the CD models as follows. Let us assign a `fugacity' $\omega$ to each domain wall, or equivalently to each cluster of pluses or minuses, in the steady state. The weight of a configuration having $S$ number of clusters is then proportional to $\omega^{S}$. We call this model the generalised coarse-grained depth (GCD) model. The CD models as defined earlier assigns equal weight to all surface configurations, which corresponds to $\omega = 1$. It turns out that the GCD model has a phase transition associated with a change in the structure of clusters, which occurs at $\omega = \omega_c = 1$.

\subsection{Phases and FDPO at the critical point}

\begin{figure}
 \begin{center}
  \includegraphics[width=11cm]{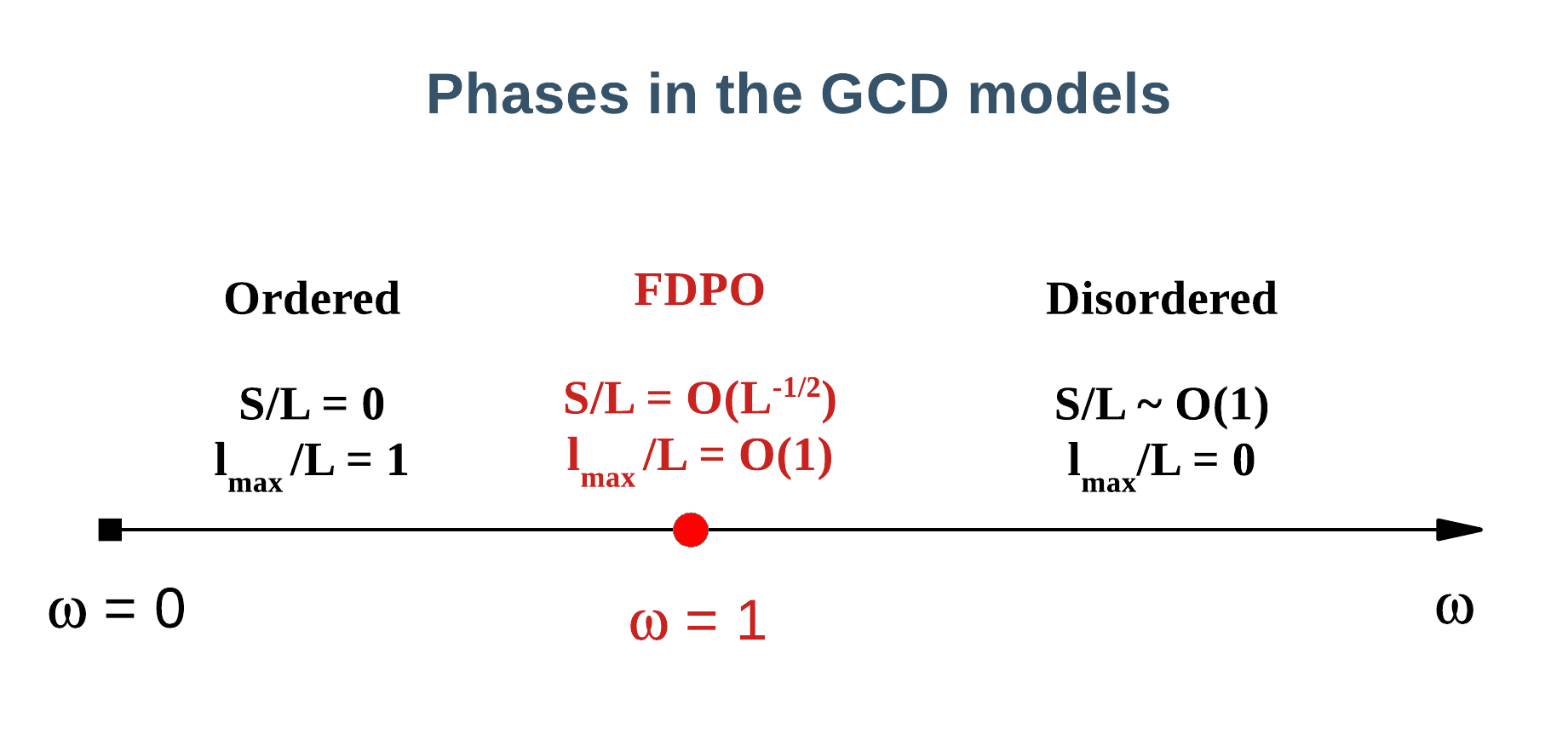}
 \end{center}
 \caption{The phase diagram of the GCD models as the cluster fugacity $\omega$ is changed: $\omega >1$ corresponds to the disordered phase and $\omega<1$ is the ordered phase. At the critical point $\omega=1$ the model becomes the regular CD model, exhibiting the FDPO behaviour. As this point the system undergoes a mixed order phase transition, as discussed in the text.}\label{phase-diag}
\end{figure}

In the GCD model the fugacity controls the typical number of clusters. For $\omega < 1$ one expects a small number of clusters and therefore a higher likelihood of having clusters of large sizes, including some of the order of system size. This promotes macroscopic order in the system. On the other hand $\omega > 1$ favours a macroscopically large number of clusters and the state is expected to be disordered; $\omega_c = 1$ is the critical point which separates these two phases. A schematic of the different phases of the CD model is shown in Figure \ref{phase-diag}.

Below we characterise the state in all the three phases for the CD2 model.

The CD2 model in steady state maps to the return problem of a random walker. Recall that the surface fluctuations of 1D EW and KPZ in the steady state are identical to that a random walk trajectory, with the spatial coordinate stands for the time and the height difference for the displacement of the walker. Then the periodic boundary condition for the fluctuating surface model in a system of size $L$ corresponds to the case where the random walker is constrained to return to the initial position at the final time $T = L$, which defines the Brownian bridge model. In the CD2 model the reference height is the height at a given site (say the site $0$), and every domain wall corresponds to the return of the random walker to the origin; equivalently one cluster in the CD2 model is the first return time in a Brownian bridge. In this section we discuss the behaviour of the system in the different phases of CD2 model with the help of the analytical results known for the return problem in a 1D random walk.

 In the GCD model, the weight of a configuration with $S$ clusters of sizes $\lbrace l_1,\cdots,l_{S} \rbrace=\lbrace 2n_1,\cdots,2n_{S} \rbrace$ in a system of size $L = 2N$ is,
\begin{equation}
 Q_L(S,\omega)=\omega^{S}\sum_{\lbrace n_1,\cdots,n_{S} \rbrace} \prod_{i=1}^{S} \frac{2^{-2n_i}}{2n_i - 1}\, ^{2n_i}C_{n_i}\,\delta_{ \sum_{i=1}^{S} n_i,N}\,, \label{weight}
\end{equation}
where we have used the statistical independence of the clusters and the known result for the probability of first return that gives the weight of a cluster of size $l_i = 2n_i$ \cite{Feller}. The partition sum for all possible configurations is,
\begin{equation}
 \Lambda_L(\omega)=\sum_{S=1}^{\infty} Q_L(S,\omega). \label{partition}
\end{equation}

We consider the grand partition function,
\begin{equation}
 \Omega_s(\omega) = \sum_{L={\rm even}} e^{-sL}\Lambda_L(\omega)=\sum_{N=1}^{\infty} e^{-2sN}\Lambda_{2N}(\omega). \label{grand}
\end{equation}
Defining $\phi(s)$ to be the generating function of the first return probability of a random walker,
\begin{equation}
 \phi(s)=\sum_{n=1}^{\infty} \frac{e^{-2ns}\,2^{-2n}}{2n-1}\, ^{2n}C_n = 1-\sqrt{1-e^{-2s}},
\end{equation}
we obtain,
\begin{equation}
 \Omega_s(\omega) = \sum_{S=1}^{\infty} \omega^{S} \phi(s)^{S} = \frac{\omega\, \phi(s)}{1-\omega\, \phi(s)}. \label{grand-phi}
\end{equation}
The partition function $\Lambda_L(\omega)$ can in principle be calculated using the Bromwich integral for the inverse Laplace transform:
\begin{equation}
 \Lambda_L(\omega) = \int_{\mathcal{B}} \frac{ds}{2\pi i}\,e^{sL}\Omega_s(\omega), \label{Bromwich}
\end{equation}
$\mathcal{B}$ is the Bromwich contour, that is a vertical contour in the complex plane with all singularities at its left. The large $L$ behaviour corresponds to small $s$, and in that limit, $\phi(s)=1-\sqrt{2s}$. Note that, $\phi(0)=1$ and $\phi(s)\le 1$ for all $s\ge 0$. Consequently, the grand partition $\Omega_s$ develops a pole at some nonzero value $s=s^*(\omega)$ for all $\omega>\omega_c=\frac{1}{\phi(0)}=1$ and the partition sum $\Lambda_L$ grows exponentially with system size. However, as $\omega \rightarrow \omega_c=1$, the pole disappears and the asymptotic behaviour of the system is controlled by the small s behaviour, $\Omega_s \sim \frac{1}{\sqrt{s}}$; $\omega_c=1$ is the critical point and the system is in a critical phase. For $\omega<1$ the grand partition is finite, signifying that only a finite number of configurations contribute nontrivially, each carrying a finite weight. This occurs when the relevant configurations carry only a handful of clusters, the large one(s) being of size of order $L$, and in fact such clusters are practically system spanning. Since the partition sum is defined over the clusters, it is straightforward to deduce the cluster size statistics from the knowledge of $\Lambda_L(\omega)$ as we discuss now. We shall follow the approach taken in \cite{trunc_ising}.

 The full distribution of a configuration having $S$ clusters with sizes $\lbrace l_1,\cdots,l_{S} \rbrace$ is given by,
\begin{equation}
 {\rm Prob.}(l_1,\cdots,l_{S})=\frac{\omega^{S}}{\Lambda_L(\omega)}\,\prod_{i=1}^{S}\,\frac{2^{-2n_i}}{2n_i - 1}\, ^{2n_i}C_{n_i}\,\delta_{ \sum_{i=1}^{S} n_i,N}, ~ n_i=l_i/2\,. \label{joint-csd}
\end{equation}
The cluster size distribution $p_L(l)\equiv {\rm Prob.}(l_1 = l; L)$ is obtained by taking the marginal over all cluster sizes except $l_1$ and over all possible number of clusters, and is given by,
\begin{eqnarray}
 p_L(l) && = \sum_{\lbrace n_2,\cdots,n_{S} \rbrace} {\rm Prob.}(l_1,\cdots,l_{S}) \nonumber\\
 && = \frac{\omega}{\Lambda_L}\,\frac{2^{-L}}{L-1}\, ^LC_{L/2}\,\delta_{l,L} + \omega \frac{2^{-l}}{l-1}\, ^lC_{l/2}\,\frac{\Lambda_{L-l}(\omega)}{\Lambda_L(\omega)}
\end{eqnarray}
where $l = 2n$. The last equality in Eq. \eqref{csd} is obtained by separating out $S = 1$ term that corresponds to the event that there is only one cluster in the entire system \cite{trunc_ising}. Using the Stirling approximation one finds that, $\frac{2^{-l}}{l-1}\, ^lC_{l/2} \approx \sqrt{\frac{2}{\pi}}\, l^{-3/2}$ for $l\gg 1$. Using this, Eqs. \eqref{weight} and \eqref{joint-csd} respectively resembles the partition sum and joint cluster size distribution of the truncated inverse-distance squared Ising (TIDSI) model, for which several analytical results for the cluster statistics are known \cite{tidsi-2014,tidsi-extr-2016, trunc_ising} and the system is known to undergo an FDPO transition \cite{trunc_ising}. 

The TIDSI model is an Ising model with a nearest neighbour interaction as well as a long range interaction which falls as $1/r^2$, but is truncated to act only within the clusters of like spins. To a good approximation, the TIDSI partition function can be written in a ``cluster representation'' which involves like spin clusters of size $\lbrace l_1,l_2,\cdots l_N\rbrace$ and takes the form \cite{tidsi-extr-2016},
\begin{equation}
Z_{\rm TIDSI}(L)=\sum_{N=1}^{\infty}\sum_{\lbrace l_1,l_2,\cdots l_N \rbrace}\prod_{n=1}^N\frac{\exp(-\beta \Delta N)}{l_n^c}\,\delta_{\sum_{n=1}^N\,l_n,\,L}. \label{tidsi}
\end{equation}
Here, $L$ is the system size, $N$ the instantaneous number of like spin clusters, $\Delta$ is an effective chemical potential for the clusters, $\beta$ is the inverse temperature, and $c=\beta\,C$ where $C$ is the strength of the long ranged interaction. The partition function of the CD2 model resembles that of the TIDSI model with $c = 3/2$ and $\e^{-\beta \Delta}=\omega/\zeta(\frac{3}{2})$, where $\zeta$ is the Reimann zeta function. The critical point $\omega_c=1$ in the CD2 model corresponds to the critical temperature in the TIDSI model, given by, $\exp(-\beta\,\Delta)_{\rm crit}=1/\zeta(\frac{3}{2})$.

In the present case, the cluster size distribution takes the form,
\begin{equation}
 p_L(l) \approx \frac{\omega}{\Lambda_L}\,\sqrt{\frac{2}{\pi}}\,L^{-3/2}\,\delta_{l,L}+\omega\,\sqrt{\frac{2}{\pi}}\,l^{-3/2}\,\frac{\Lambda_{L-l}(\omega)}{\Lambda_L(\omega)}. \label{csd}
\end{equation}
Now we quantify the partition sum $\Lambda_L(\omega)$ and the cluster statistics in the three phases.\\

{\bf Disordered phase $(\omega>1)$:} In this case the grand partition in Eq. \eqref{grand-phi} has a simple pole at $s = s^*$ such that $\phi(s^*) = 1/\omega$, or,
\begin{equation}
s^*=s^*(\omega)= - \frac{1}{2} \ln\left[1- \left(1-\frac{1}{\omega} \right)^2 \right]>0, \label{pole}
\end{equation}
using the expression for $\phi(s)$ and solving for the pole. Subsequently, from Eq. \eqref{Bromwich}, the partition function is obtained as,
\begin{equation}
 \Lambda_L(\omega)=\frac{e^{s^*\,L}}{\omega\,\phi'(s^*)},~\phi'(s^*)=1-(1-1/\omega)^2. \label{disordered}
\end{equation}
To find the cluster size statistics, we note from Eq. \eqref{csd} that the first term (single system spanning cluster) is exponentially small in the system size, and the distribution is,
\begin{equation}
 p_L(l) \approx \omega\,\sqrt{\frac{2}{\pi}}\,l^{-3/2}\,e^{-s^*\,l}, \label{disordered-csd}
\end{equation}
which implies that the typical clusters are finite (of sizes $1/s^*(\omega)$), the state is disordered with macroscopic number of clusters ($S\sim \mathcal{O}(L)$), and the largest cluster size is $\sim \ln(L)$.\\

{\bf Critical phase $(\omega=1)$:} This is the `classic' CD model. As we approach criticality $\omega \rightarrow \omega_c=1+,~s^*\rightarrow 0$ implying that the exponential form of the partition function no longer holds and the typical cluster sizes diverge. In this limit the grand partition in Eq. \eqref{grand-phi} takes the form, $\Omega_s=\frac{\omega\, \phi(s)}{1-\omega\, \phi(s)}\rightarrow \frac{1}{\sqrt{s}}$. Consequently, in the large $L$ limit,
\begin{equation}
 \Lambda_L(\omega_c)= \frac{1}{\sqrt{2\pi}}\,L^{-1/2}. \label{critical}
\end{equation}
From Eq. \eqref{csd} one finds the the first term in the cluster size distribution falls as $L^{-1}$ and have negligible contribution to the distribution and the moments. The dominant contribution comes from the second term, which gives, in the limit when $L, l, L - l \gg 1$, $ p_L(l)\approx \omega \sqrt{\frac{2}{\pi}}\,l^{-3/2}\,(1-l/L)^{-1/2},$
which, for $l\ll L$, takes the form,
\begin{equation}
 p_L(l)\approx \omega \sqrt{\frac{2}{\pi}}\,\,l^{-3/2}. \label{critical-csd-l}
\end{equation}
It is important to note that, here the mean cluster size indeed diverges, but only as $L^{1/2}$ implying that the typical number of clusters $S \sim O(\sqrt{L})$; the state, however, always carries macroscopic clusters of size $\sim L$ \cite{MB_fdpo}.
For the TIDSI model at the critical point with $1 < c < 2$, the largest cluster is known to be macroscopic \cite{tidsi-extr-2016}; in fact it was found that the average size of the $k$'th largest cluster for any finite $k$ is proportional to $L$. As mentioned earlier, the CD model corresponds to $c = 3/2$ in the TIDSI model and therefore carries the same feature. The critical state is ordered, and the existence of large number of macroscopic clusters implies the existence of macroscopic fluctuations, indicating that system is in FDPO state at criticality. Two point correlation results at the critical point of TIDSI model \cite{trunc_ising}, which could be directly extended in the present case, also corroborate the same conclusion.\\

{\bf Ordered phase ($\omega<1$):} Here the leading term of the grand partition is a constant, and the singular behaviour is captured in the leading s-dependent term as,
\begin{equation}
 \Omega_s(\omega)\approx \frac{\omega}{1-\omega} \left(1-\frac{\sqrt{2s}}{1-\omega} \right).
\end{equation}
This gives the leading large $L$ behaviour of the partition sum to be
\begin{equation}
  \Lambda_L(\omega)= \frac{\omega}{(1-\omega)^2}\,L^{-3/2}. \label{ordered}
\end{equation}
The cluster size distribution in this phase is given by
\begin{equation}
 p_L(l)\approx \sqrt{\frac{2}{\pi}}\,[(1-\omega)^2\,\delta_{l,L}+\omega\,l^{-3/2}\,(1-l/L)^{-3/2}]. \label{ordered-csd}
\end{equation}
Note that here a single system-spanning cluster appears with a finite probability. The mean cluster size is $\langle l\rangle \sim L$ to the leading order, which immediately tells that, in all the accessible configurations, the number of clusters is finite ($S \sim \mathcal{O}(1)$). The correspondence to TIDSI model with $c = 3/2$ asserts that the state appears with only one macroscopic cluster that covers almost the entire system, $\langle l_{\rm max}\rangle \sim L - \mathcal{O}(\sqrt{L})$ \cite{tidsi-extr-2016}. Therefore, the ordered phase either appears with a completely ordered state with a single cluster of size exactly equal to $L$, or there is one large nearly system-spanning cluster along with a finite number of small clusters of sizes not larger than $\mathcal{O}(\sqrt{L})$.

\subsection{Mixed order transition at $\omega=1$}
At the critical point, the onset of order is identified by observing the emergence of a macroscopic cluster. An appropriate quantity to characterise the state of order in all three phases is the average largest cluster size, $x = \langle l_{\rm max}\rangle/L$. As $\omega$ is increased from $0$ to $1$, $x = 1$. At the critical point $x$ takes a finite value \cite{tidsi-extr-2016}. For all values of $\omega > 1$, we have $x \propto \ln L/L \rightarrow 0$, marking a discontinuous transition, from $0$ to a finite value. It is instructive to look at the correlation of fluctuations $G(r) = \langle \sigma_i \sigma_{i+r}\rangle$ near the critical point. Exactly at criticality the order is long-ranged, as already reported for the CD models \cite{MB_fdpo} as well as for the TIDSI model \cite{trunc_ising}. In the disordered phase, the correlation is exponentially decaying, with the correlation length governed by the typical cluster sizes, $\xi \propto 1/s^*(\omega)$. However, as we approach the critical point from the disordered phase $(\omega = 1+)$, the correlation length diverges, which is typical for a continuous phase transition. The co-occurence of diverging correlation length and a discontinuous ordering field indicates a mixed order transition.

\section{Concluding remarks}
The coarse-grained depth (CD) model is one of the simplest models of fluctuation dominated phase ordering (FDPO), amenable to detailed analytical and numerical studies. This model is also related to other models like the sliding particle (SP) and truncated inverse distance squared Ising (TIDSI) models which show FDPO behaviour. It should be noted that the interactions in the CD model are effectively long ranged in the sense that a single flip of a hill or a valley of the underlying surface can cause substantial reconfiguration of the domain walls in the entire system. An alternative manifestation of this is through the mapping to the TIDSI model, in which spin-spin interactions have a long range. Thus the present work provides new results for coarsening in long-ranged interacting systems, which are relatively less studied but can have interesting features \cite{Corberi2021}.
The method and results of the study in this paper bring out not only interesting nontrivial features of the model itself and FDPO, but also help to identify features that are expected to be more generic, applicable to coarsening systems in general.

In this work we argued that the CD models can be characterised using the number of domain walls $S$ and the maximum cluster size $l_{\rm max}$ both in the steady state and in the coarsening regime. We defined a generalised CD (GCD) model with a chemical potential-like parameter $\omega$ assigned to each cluster. The GCD models exhibit a mixed order phase transition from disordered to ordered phase, on tuning $\omega$. FDPO occurs at the critical point $\omega=\omega_c = 1$ that corresponds to the regular CD models, implying that it is a critical phase. However, FDPO is very different from a regular critical phase in that, unlike the latter, in FDPO the steady state two-point correlation $G(r)$ involves the system size $L$ in a crucial way -- in fact it is a function of $r/L$ \cite{MB_fdpo}, rather than $r$ alone, as $L \rightarrow \infty$. At $\omega=\omega_c$ the distribution of $S$ is broad and scales with $\sqrt{L}$, signifying the presence of macroscopic domain wall structures, leading to the violation of Porod law. On the other hand the steady state distribution of $l_{\rm max}$ scales with the system size $L$, and manifests macroscopic fluctuations of the ordered domains. Note that $\langle l_{\rm max}\rangle \sim \mathcal{O}(L)$ in the ordered phase too, but in this phase $\langle S \rangle ~{\rm is~ of}~ \mathcal{O}(1)$; whereas in the disordered phase these two quantities take values $\sim \mathcal{O}(\ln L)$ and $\mathcal{O}(L)$ respectively. Although the results are obtained for a simple class of models, the pair $(S,l_{\rm max})$ may help characterise FDPO and other phases in other models as well, where ordered regions are associated with an uninterrupted cluster of a single species.

 An important feature observed in the present work is the existence of strong corrections ($\sim d/t^{\theta}$) to the leading power law behaviour of the coarsening length $\mathcal{L}(t)$, given in Eq. \eqref{Lt-correction}. 
We found that the scaling of the correlation $G(r,t)$ and the number density of clusters $s(t)$ both depend sensitively on this feature. We evaluated the exponent $\theta$ by noting that, this correction gives rise to a subleading time dependent term in the average maximum cluster size $\langle l_{\rm max}(t)\rangle$. When this term, evaluated at times $t\sim L^z$, is compared with the observed subleading behaviour of $\langle l_{\rm max}\rangle$ in the steady state, one finds $\theta=1/2z$.

During coarsening, the behaviour of the CD models within a growing length scale $\mathcal{L}(t)$ resembles the steady state behaviour. Thus, anomalously large and growing fluctuations govern the state of the system while coarsening \cite{Iyer22}. However, as we have shown in section \ref{coarsening-cds}, the variance of $s(t)$ in the coarsening regime is time independent. Further, in the mean largest cluster size $\langle l_{\rm max}(t)\rangle$ there are multiplicative logarithms in $t$ and $L$ to the leading order, thereby strongly modifying the expected power law growth. The latter behaviour has its origin in the extreme value statistics: the largest cluster size can be thought of as the largest of $N_{\rm sub}(t)=L/\mathcal{L}(t)$ number of locally largest and almost independent clusters sampled from the tail of the cluster size distribution, which in the CD3 model falls exponentially. Consequently $l_{\rm max}(t)$ follows a Gumbel distribution with mean proportional to $(a_1+g_1\,\ln (N_{\rm sub}(t)))$.

We may expect that such power law corrections at large but finite times, as well as the multiplicative logarithms in the extremal quantities, may arise in other coarsening systems as well.

Finally, an in depth understanding of the scaling behaviour, in particular, the power law corrections in coarsening is of general importance. To this end, a renormalisation group approach might be useful \cite{Bray94, Bray90}. Such an approach may also give deeper insights into the nature of FDPO. Further, the detailed fractal structure of the domains in the CD model and other models showing the FDPO behaviour is an interesting question. Systematic studies in these directions would be worthwhile.

\section*{Acknowledgement}
The authors acknowledge useful discussions with Eli Barkai, Satya N Majumdar, Saroj Nandi, Souvik Sadhukhan and Gr\'egory Schehr.
This project was funded by intramural funds at TIFR Hyderabad from the Department of Atomic Energy (DAE), India.
M.B. acknowledges support under the DAE Homi Bhabha Chair Professorship of the Department of Atomic Energy, India. He acknowledges the Erwin Schr\"odinger Institute (ESI) for support during the thematic programme `Large Deviations, Extremes and Anomalous Transport in Non-equilibrium Systems', and the International Centre for Theoretical Sciences (ICTS-TIFR) for support during the discussion meeting `Statistical Physics of Complex Systems'.

\end{document}